\documentclass[letterpaper]{article} 
\usepackage{aaai24}  
\def\mytitle{Removing Interference and Recovering Content Imaginatively for Visible Watermark Removal}

\usepackage{times}  
\usepackage{helvet}  
\usepackage{courier}  
\usepackage[hyphens]{url}  
\usepackage{graphicx} 
\usepackage{natbib}  
\usepackage{caption} 
\usepackage{algorithm}
\usepackage{algorithmic}

\usepackage{newfloat}
\usepackage{listings}
\DeclareCaptionStyle{ruled}{labelfont=normalfont,labelsep=colon,strut=off} 
\floatstyle{ruled}
\newfloat{listing}{tb}{lst}{}
\floatname{listing}{Listing}
\title{\mytitle}
\author{
    Yicheng Leng\textsuperscript{\rm 1, \rm 2}, Chaowei Fang\textsuperscript{\rm 1}\thanks{Corresponding author.}, Gen Li\textsuperscript{\rm 1, \rm 3}, Yixiang Fang\textsuperscript{\rm 2}, Guanbin Li\textsuperscript{\rm 4, \rm 5}
}
\affiliations{
    \textsuperscript{\rm 1} School of Artificial Intelligence, Xidian University, Xi’an, China\\
    \textsuperscript{\rm 2} School of Data Science, The Chinese University of Hong Kong, Shenzhen, China\\
    \textsuperscript{\rm 3} Afirstsoft, Shenzhen, China\\
    \textsuperscript{\rm 4}  School of Computer Science and Engineering, Research Institute of Sun Yat-sen University in Shenzhen, Sun Yat-sen University, Guangzhou, China\\
    \textsuperscript{\rm 5} GuangDong Province Key Laboratory of Information Security Technology\\
}

\usepackage{epsfig}
\usepackage{amsmath}
\usepackage{dsfont}
\usepackage[super]{nth}
\usepackage{multirow}
\usepackage{color}
\usepackage{amsfonts} 
\usepackage{booktabs}

\definecolor{DeltaColor}{rgb}{0.039,0.73,0.71}
\definecolor{SetaColor}{rgb}{0.867, 0.0235, 0.376}
\definecolor{SigmaColor}{rgb}{0.98,0.45,0.0}
\definecolor{RedColor}{rgb}{0.8,0,0}
\definecolor{AlphaColor}{rgb}{0,0,0.8}
\definecolor{BetaColor}{rgb}{0.8,0,0.8}
\definecolor{GammaColor}{rgb}{0.5,0,0.7}
\definecolor{EpsilonColor}{rgb}{0.353,0.725,0.906}
\definecolor{TauColor}{rgb}{0.423,0.235,0.192}
\definecolor{WtColor}{rgb}{0.235,0.470,0.470}

\graphicspath{{./figures/}}

\begin{document}

\maketitle
\begin{abstract}
Visible watermarks, while instrumental in protecting image copyrights, frequently distort the underlying content, complicating tasks like scene interpretation and image editing. Visible watermark removal aims to eliminate the interference of watermarks and restore the background content. However, existing methods often implement watermark component removal and background restoration tasks within a singular branch, leading to residual watermarks in the predictions and ignoring cases where watermarks heavily obscure the background. To address these limitations, this study introduces the \textit{Removing Interference and Recovering Content Imaginatively} (RIRCI) framework. RIRCI embodies a two-stage approach: the initial phase centers on discerning and segregating the watermark component, while the subsequent phase focuses on background content restoration. To achieve meticulous background restoration, our proposed model employs a dual-path network capable of fully exploring the intrinsic background information beneath semi-transparent watermarks and peripheral contextual information from unaffected regions. Moreover,  a \textit{Global and Local Context Interaction} module is built upon multi-layer perceptrons and bidirectional feature transformation for comprehensive representation modeling in the background restoration phase. The efficacy of our approach is empirically validated across two large-scale datasets, and our findings reveal a marked enhancement over existing watermark removal techniques.
\end{abstract}
\section{Introduction}
Visible watermarks serve to safeguard image copyrights. Despite their utility, these watermarks introduce significant interferences to background images, thereby hampering tasks such as scene interpretation and image editing. Consequently, the removal of visible watermarks to recover the background has emerged as a research area of paramount importance. Furthermore, examining the robustness of these visible watermarks against adversarial attacks is also imperative. This work specifically addresses the removal of watermarks characterized by varying transparency and the restoration of the background.

\begin{figure}[t]
\begin{center}
    \includegraphics[width=\linewidth]{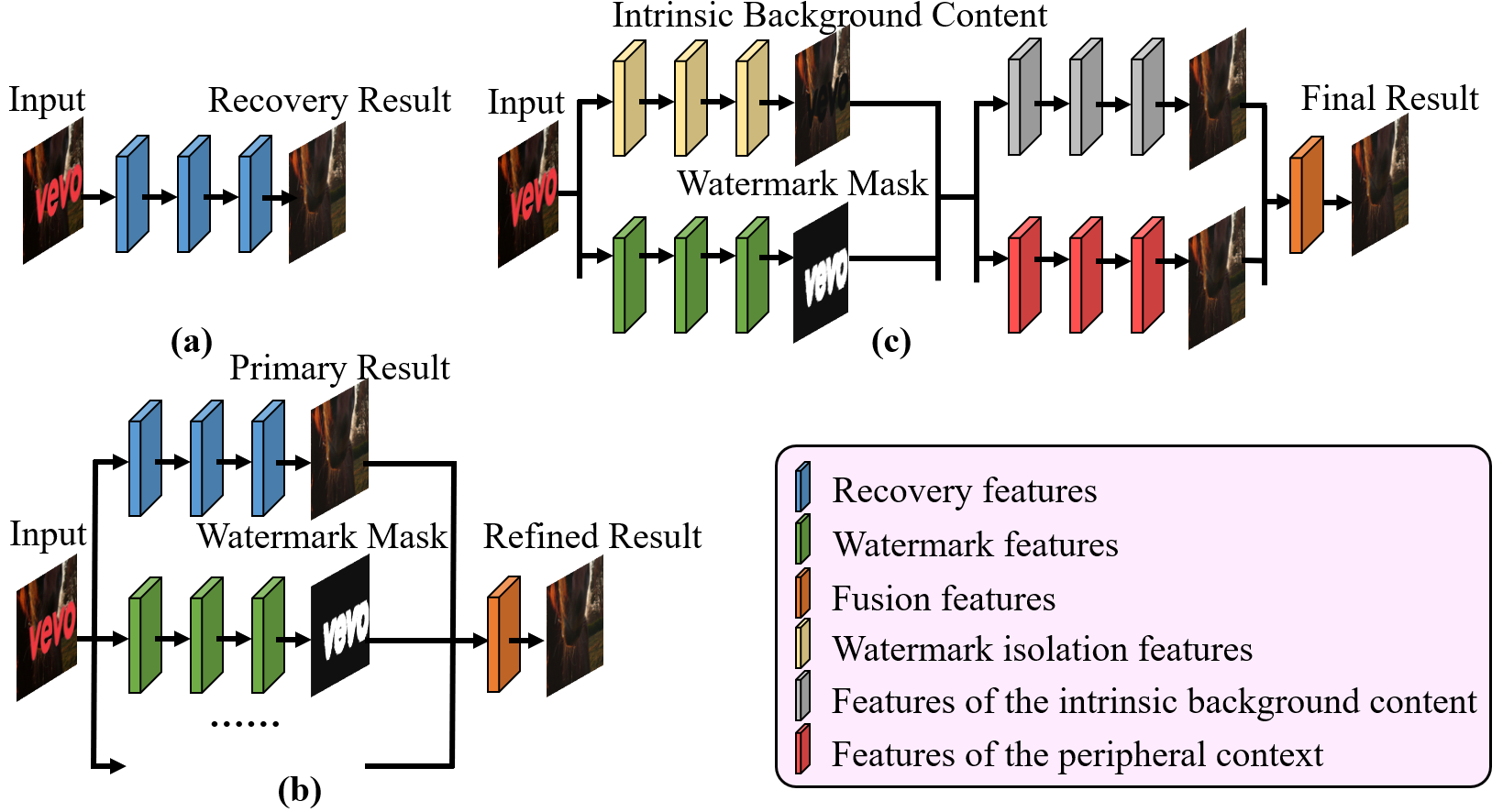}
\end{center}
   \caption{Overview of different watermark removal frameworks: (a) Direct image-to-image transform; (b) Multi-task learning; (c) Ours.
   }
\label{fig0}
\end{figure}

The chief challenges revolve around the complete elimination of watermark components and the precise recovery of the background. Existing watermark removal methodologies can be categorized into two types. Early approaches view the task as a direct image-to-image transformation, employing convolutional neural networks to deduce the background~\cite{cheng2018large,li2019towards}, as depicted in Fig.~\ref{fig0}~(a). Others leverage multi-task learning to concurrently pinpoint watermark locations and restore the background, as depicted in Fig.~\ref{fig0}~(b). Notably, \citet{liu2021wdnet} utilize an encoder-decoder network for predicting watermark mask, transparency, and image. \citet{hertz2019blind,cun2021split,liang2021visible} employ dedicated decoding branches to accomplish different tasks. Given that watermark contents remain orthogonal to the background, achieving the dual objectives of watermark removal and background recovery necessitates distinctive feature representations. However, existing methods often amalgamate the two sub-tasks within a singular branch, resulting in residual watermarks in the predictions. 
Moreover, the peripheral contextual information from unaffected regions surrounding watermarks is paramount for background recovery in extenuating circumstances where a watermark severely obscures background references.
This is not specifically taken into consideration by existing methods.

\begin{figure}[t]
\begin{center}
    \includegraphics[width=\linewidth]{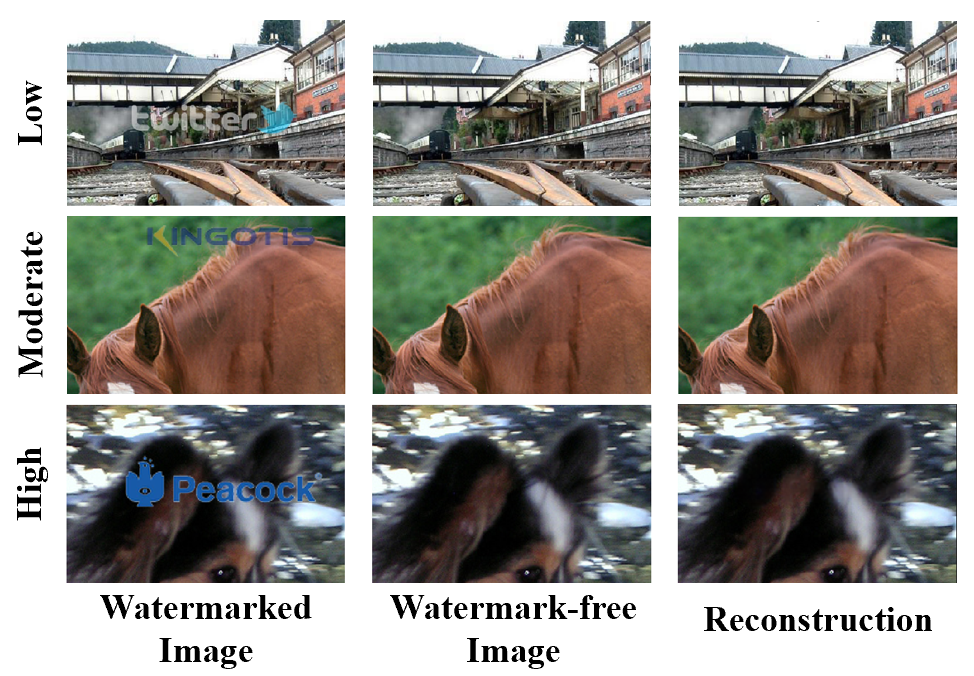}
\end{center}
   \caption{Illustration of our method's ability in removing watermarks with low, moderate, and high opaqueness.
   }
\label{fig1}
\end{figure}

In response to the aforementioned challenges, this study presents a novel two-stage framework, named \textit{Removing Interference and Recovering Content Imaginatively} (RIRCI), as depicted in Fig.~\ref{fig0}~(c). Within RIRCI, the task of watermark component removal is distinctly partitioned from that of background restoration. 
The primary stage is dedicated to pinpointing the watermark's spatial region and subsequently isolating its component, which also reveals the intrinsic background content.
The subsequent stage aims to restore the obscured background information, a consequence of diminished luminance or occlusional interference.

The restoration process hinges on a dual informational paradigm: the intrinsic background information visible beneath semi-transparent watermarks and the peripheral contextual data from undisturbed regions surrounding the watermark. 
To harness these informational facets, there's a necessity for a holistic model discerning both macroscopic and microscopic contextual nuances. Addressing this requirement, we devise a \textit{Global and Local Context Interaction} module, on the basis of multi-layer perceptrons with variant receptive fields and bidirectional feature transformation mechanism.
Utilizing the aforementioned module as a foundational block, we construct a dual-path network dedicated to the task of background restoration. The initial pathway is devoted to recovering the image leveraging the intrinsic background content. The secondary pathway, taking cues from an image inpainting aglorithm~\cite{suvorov2022resolution}, innovatively reconstructs the watermark-distorted region, grounding its logic in the peripheral unaffected background context. 
The final outcome of the restoration process is derived from the fusion of both pathways. Fig.~\ref{fig1} demonstrates the ability of our method in removing watermarks with various opaqueness levels.
Our empirical validations, spanning two large-scale datasets, affirm the superiority of our approach against contemporary methodologies.

Main contributions of this manuscript include:
\begin{itemize}
    \item We introduce a novel two-stage framework by decomposing the visible watermark removal task into discrete tasks of watermark component exclusion and background content restoration.
    \item We build up a dual-path background content restoration network by exploring both intrinsic background content inside the watermark-distorted region and peripheral context information from the unaffected region. 
    \item We propose a global and local context interaction module, tailored for exhaustive feature representation conducive to the background restoration process.
    \item We conduct extensive experiments on two large-scale datasets, and the results indicate that our method significantly outperforms existing state-of-the-art methods.
\end{itemize} 
\section{Related Work}
\subsection{Visible Watermark Removal} 
The objective is to transform watermark-distorted images into their watermark-free counterparts. \citet{cheng2018large} pioneer the deployment of deep convolutional neural networks for this task through image-to-image transformation. Subsequent works, such as \cite{li2019towards,cao2019generative}, employ generative adversarial learning to enhance the authenticity of the recovered images. Nevertheless, these techniques have difficulty in coping with watermarks exhibiting diverse attributes like opacity, hue, and form.
\begin{figure*}[t]
\begin{center}
    \includegraphics[width=1\linewidth]{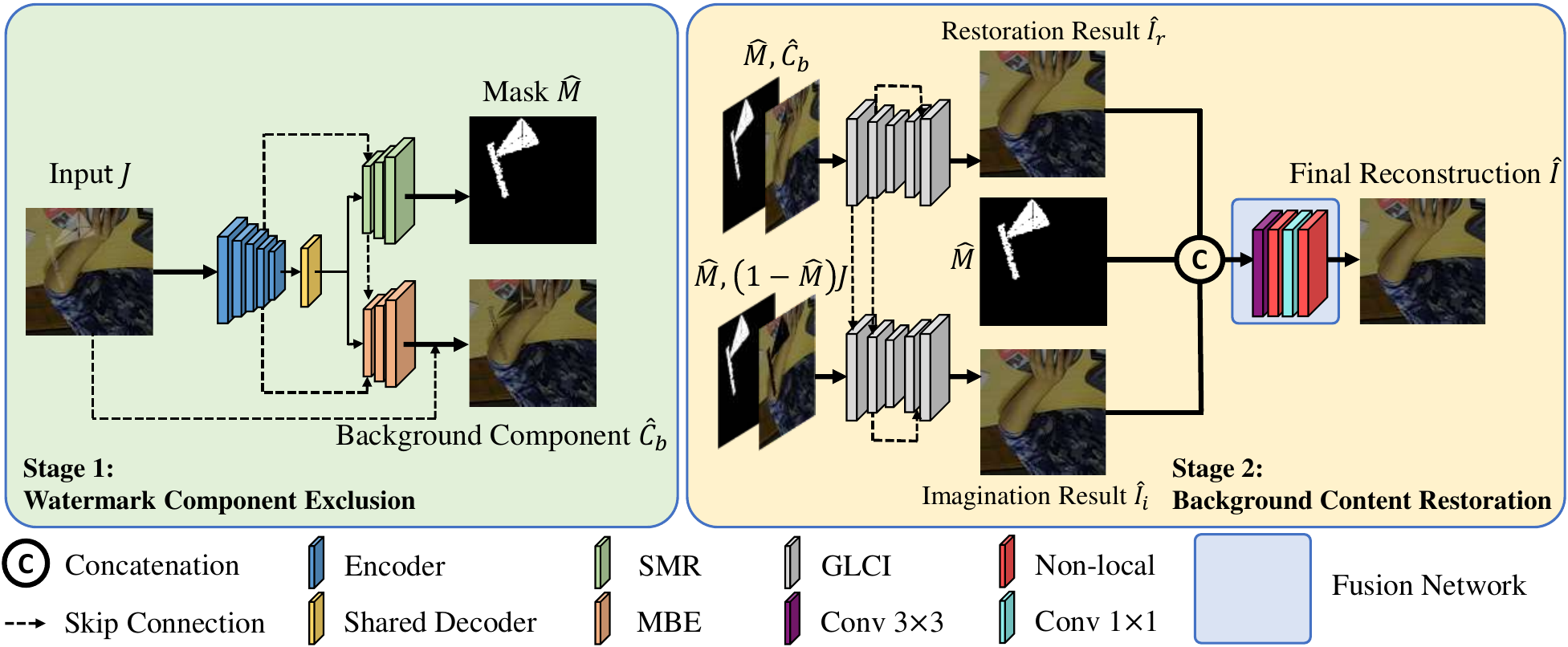}
\end{center}
   \caption{Our proposed visible watermark removal method is composed of two phases. The first phase employs a U-shape model with dual decoding branches for predicting watermark mask and excluding watermark component respectively. The second phase concentrates on reconstructing the background image from two aspects: recovering the background image from the intrinsic background component and filling the watermark region according to peripheral contextual information. The two reconstructed background images are fused to derive the final output.
   }
\label{fig3}
\end{figure*}
Contemporary solutions leverage multi-task learning for visible watermark removal. \citet{hertz2019blind} introduce a model with distinct decoding branches to separately deduce the background, motif mask, and motif image. The motif's removal is facilitated by compositing the inferred background with the input, guided by the motif mask. Following a similar multi-task paradigm, \citet{cun2021split} incorporate the channel attention mechanism for specialized feature extraction across sub-tasks. \citet{liang2021visible} refine mask predictions via a coarse-to-fine strategy, utilizing the resultant masks to enhance image reconstruction features. \citet{sun2023denet} endeavor to disentangle backgrounds from watermarks in an advanced embedding space, while \citet{liu2021wdnet} tackle the watermark removal challenge by inferring  mask, opacity, and color of watermarks. Notably, aforementioned methods~\cite{cun2021split,liang2021visible,liu2021wdnet,sun2023denet} invariably integrates a refinement phase for background reconstruction.

A prevailing oversight in existing methods is the neglect of the watermark's inherent orthogonality to the background. They often combine the processes of watermark component exclusion and background restoration within a singular inferential domain. Addressing this issue, our research presents an innovative two-stage framework, distinctly separating watermark component exclusion from background restoration. Additionally, we recognize and remedy the previous methods' deficiency in harnessing context from undistorted zones. Our dual-path restoration model can navigate both intrinsic background data and contextual cues from undistorted zones.

\subsection{Image Inpainting} 
Image inpainting endeavors to reconstruct absent regions in images by leveraging contextual information from their surroundings. This domain garners significant scholarly attention~\cite{pathak2016context,dong2022incremental,li2022mat,suvorov2022resolution,liu2023coordfill}. Through meticulous architectural designs and optimization strategies, these methodologies can predict plausible content for the void areas, evidencing their proficiency in extracting the surrounding contextual features. Such proficiency in contextual feature extraction holds potential advantages for the task of visible watermark removal, particularly when watermarks severely obscure the background. However, a mere adaptation of image inpainting approaches to watermark removal risks overlooking the intrinsic background content residing beneath semi-transparent watermarks. To address these issues, we introduce a sophisticated dual-path model for background restoration, taking advantage of both the intrinsic background content and contextual cues from watermark-unaffected zones.

\section{Methodology}

\subsection{Problem Definition}
This paper is targeted at transforming a watermarked image $J \in \mathbb R^{h\times w\times 3}$ into its undistorted version $I \in \mathbb R^{h\times w\times 3}$, where $h$ and $w$ represents image height and width, respectively.
Fundamentally, the formation of a watermarked image can be interpreted as below:
\begin{equation}\label{eq1}
J = A\circ W + (1 - A)\circ I,
\end{equation}
where $W\in \mathbb R^{h\times w\times 3}$ represents the watermark image, and $A\in [0,1]^{h\times w\times 1}$ denotes the opaqueness channel. We denote the region contaminated by the watermark image as $M$, where $M=A>0$.
From the above formulation, $J$ can be regarded as the addition of two components, including the watermark component denoted as $C_w=A\circ W$, and the intrinsic background component denoted as $C_b=(1 - A)\circ I$.

\subsection{Approach Overview}
We decompose the above visible watermark removal into two sub-tasks: watermark component exclusion and background content restoration.
As shown in Fig.~\ref{fig3}, a two-phase framework is built up to tackle them separately, considering the information of the watermark and background is orthogonal to each other. The first phase focuses on locating the watermark region and extracting out the watermark component; the second phase aims to restore the background content using both the intrinsic background information in the watermark region and the peripheral contextual information in the watermark-unaffected region. Methodology details of two phases are illustrated below.

\subsection{Watermark Component Exclusion}
This phase contributes to locating the region distorted by watermark and excluding the watermark component, serving as the foundamental brick for the subsequent background restoration bricks.
We adopt the U-shape convolutional neural network (CNN) in~\cite{liang2021visible} to implement watermark component exclusion.
The encoder part of the CNN model is composed of five stages built with conventional convolution layers and residual blocks. Each stage decreases the spatial dimensions of feature maps by half.
The decoder part is consisting of a shared decoding block and dual branches tailored for inferring the watermark mask and component, respectively.
Here, the shared decoding block comprises of residual blocks as well. The watermark mask inference branch is built with a series self-calibrated mask refinement modules, while the watermark component inference branch is enhanced with predicted watermark masks. 
Given the input image $J$, we denote the predicted watermark mask as $\hat M$ and watermark component as $\hat C_w$. 
The intrinsic background component $\hat C_b$ can be obtained by subtracting $\hat C_w$ from $J$, i.e. $\hat C_b=J-\hat M \circ \hat C_w$. 

\subsection{Background Content Restoration}
This phase is targeted at restoring the background content with reference information from intrinsic background component beneath the semi-transparent watermark and unaffected regions. To resolve this problem, we construct dual pathways to make full usage of the two kinds of reference information respectively. Moreover, to develop comprehensive feature extraction mechanism, we introduce a global and local context interaction module as the basic blocks of the background restoration model.

\subsubsection{Global and Local Context Interaction Module}

\begin{figure}[t]
\begin{center}
    \includegraphics[width=\linewidth]{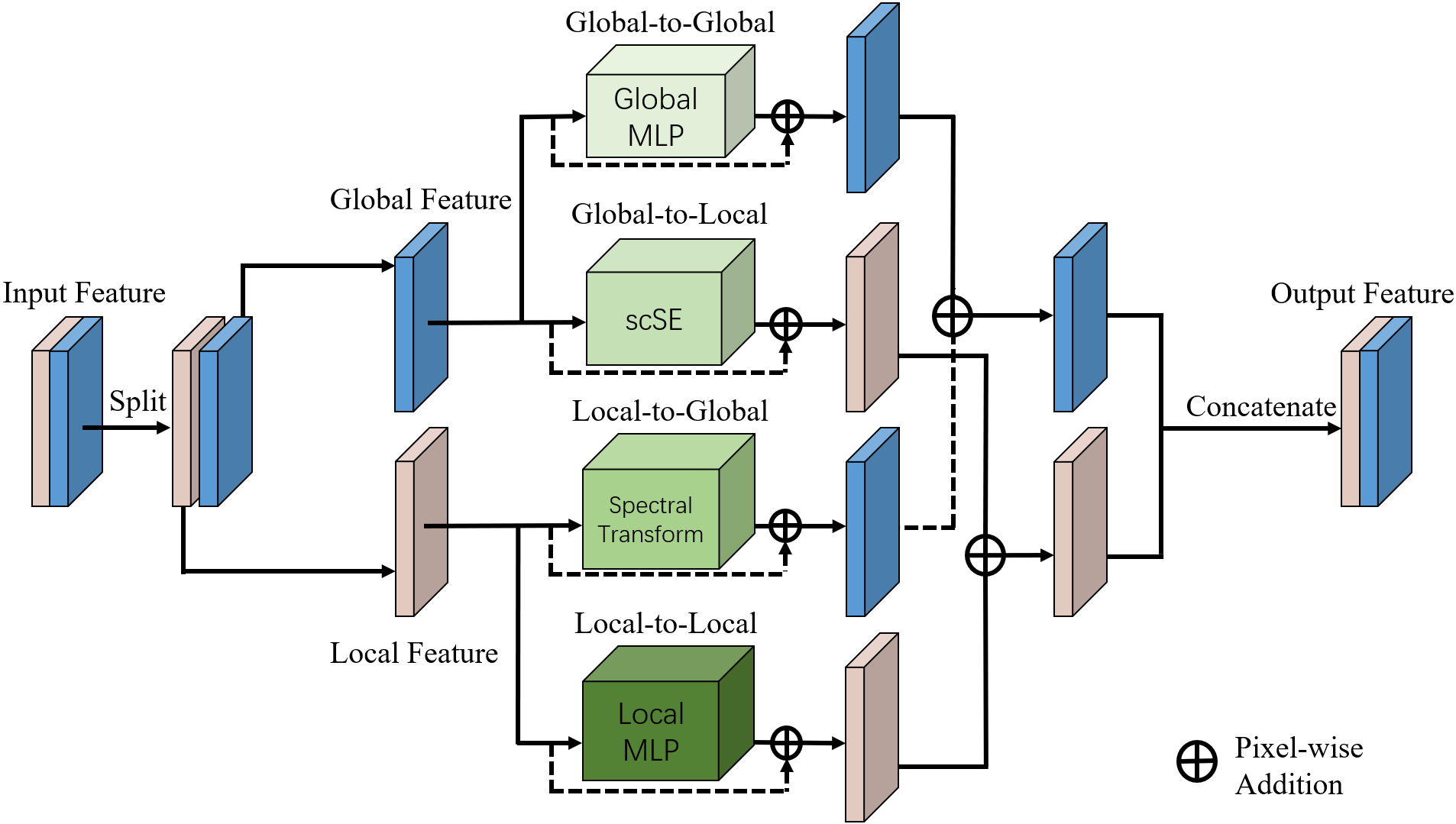}
\end{center}
   \caption{Illustration of the global and local context interaction module. Global and local MLP branches~\cite{tu2022maxim} are used for extracting global and local features, respectively. The information propagation from global to local branch is implemented with the spatial an channel squeeze and excitation block~\cite{roy2018concurrent}, while that from local to global branch is implemented with the spectral transform module~\cite{chi2020fast}.  }
\label{fig4}
\end{figure}

To cope with diverse and variably transparent watermarks, our restoration model needs to effectively leverage information both within and outside the watermark region in the image, which is achieved by extracted local and global feature maps, respectively. The utilization of global features is crucial for reconstructing broad structural contexts and extensively distorted areas; conversely, the integration of local features is pivotal for the restoration of fine details and moderately distorted areas. Therefore, we design \textit{Global and Local Context Interaction} (GLCI) module, which enables concurrent extraction of local and global features through a interactive learning approach, as depicted in Fig. \ref{fig4}.

\begin{figure}[h]
\begin{center}
    \includegraphics[width=0.9\linewidth]{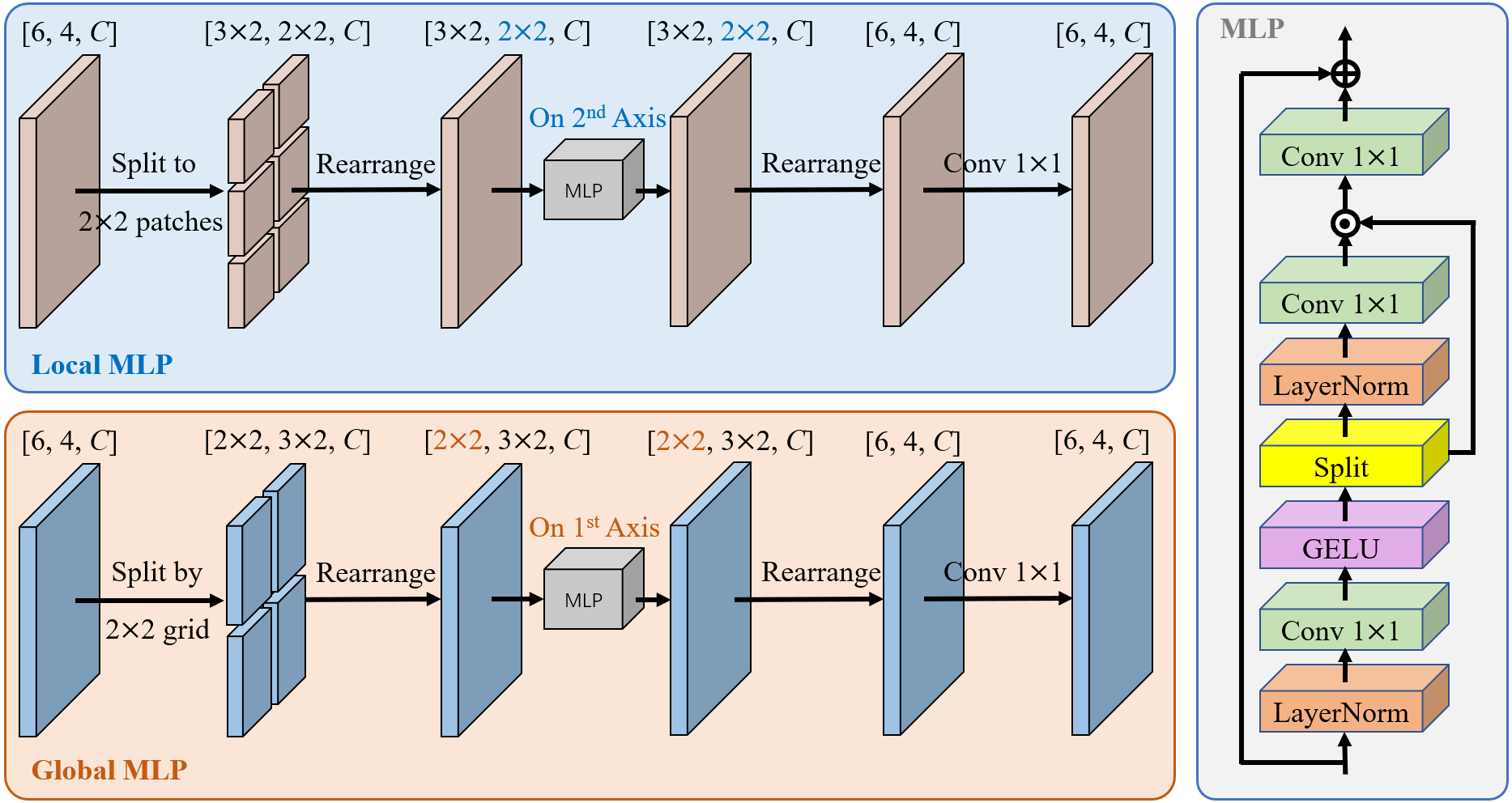}
\end{center}
   \caption{Illustration of local and global MLP. Suppose the input feature has size of $6\times4$. The local MLP split the feature into 6 patches with size of 2$\times$2. In the calculation process of MLP, the senocd axis is regarded as the channel dimension.
   In contrast, the global MLP split the feature into 4 patches with size of $3\times 2$, and the first axis is regarded as the channel dimension.}
\label{fig5}
\end{figure}

Local features are acquired through the modelling of inter-pixel correlations within a designated local region, while global features stem from exploring pixel dependencies spanning different local regions. Therefore, we introduce the local multi-layer perceptron (MLP) and global MLP for feature extraction, inspired by \cite{tu2022maxim}. These elements first divide the feature map into uniformly sized patches. By utilizing distinct axes as channels, convolutional layers are then employed to investigate relationships within individual patches or across divergent patches. 
The resulted feature patches are subsequently re-organized into a complete feature map.
An example is depicted in Fig.~\ref{fig5}.

The above local and global feature extraction modules may fall short due to inherent limitations in local and global information. 
For example, with highly opaque watermarks, local features need be inferred from global information. Conversely, substantial and transparent watermarks necessitate extracting global features from abundant yet concealed local information. To address this, we integrate reciprocal local-global feature propagation mechanisms within the GLCI framework. Specifically, we employ a spectral transform block~\cite{chi2020fast} to derive global features from the frequency domain of local features. This involves utilizing the Fast Fourier Transform (FFT) to explore context information from the entire feature map. 
Simultaneously, a spatial and channel squeeze and excitation block (scSE) \cite{roy2018concurrent} is utilized to identify pertinent global information through attention computation, thereby enhancing the local features.

\subsubsection{Dual-path Background Restoration Model}
Leveraging GLCI as the fundamental module, we build up a dual-path background restoration model composed of a content restoration sub-network and a content imagination sub-network. 
The content restoration sub-network is targeted at recovering the background image from the background component, It regards the concatenation of $\hat M$ and $\hat C_b$ as the inputs, deriving $\hat I_r$. 
High watermark opacity can limit the informativeness of the background component within the watermark-affected region. Hence, we introduce the content imagination sub-network which regards the masked image $(1-\hat{M})J$ and $\hat{M}$ as inputs and outputs $I_i$.
Finally, $\hat I_r$, $I_i$ and $\hat M$ are fed into a fusion module built upon non-local blocks~\cite{wang2018non}, producing the ultimate result $\hat I$. 

\subsection{Objective Function}
We employ $L_1$ loss and perceptual loss to regularize network outcomes: $\hat{C_b}$, $\hat{I_r}$, $\hat{I_i}$ and $\hat{I}$.
Given an predicted image $X$ and its ground-truth image $Y$, the $L_1$ loss is calculated as follows,
\begin{equation}
    \ell_1(X, Y) = ||X-Y||_1.
\end{equation}
For highlighting the reconstruction of watermark-affected region, we also introduce a masked $L_1$ loss:
\begin{equation}
    \ell_1^{msk}(X, Y, M) = ||M\circ(X-Y)||_1.
\end{equation}
We extract features with the VGG16 model~\cite{simonyan2014very} pretrained on ImageNet~\cite{deng2009imagenet} to calculate the perceptual loss:
\begin{equation}
    \ell^{vgg}(X, Y) =  \sum_{k=1}^3 ||\Phi_{vgg}^{(k)}(X) - \Phi_{vgg}^{(k)}(Y)||_1.
\end{equation}
The training losses for $\hat{C_b}$, $\hat{I_r}$, $\hat{I_i}$ and $\hat{I}$ are formed by combining the above loss terms:
\begin{align}
    L_b = & \lambda_1 \ell_1^{msk}(\hat C_b, C_b, M) + \lambda_2 \ell^{vgg}(\hat C_b, C_b) \\
    L_r = & \lambda_1 [\ell_1^{msk}(\hat I_r, I_r, M\circ(A>\alpha)) + \gamma \ell_1(\hat I_r, I_r)] \nonumber \\
    & + \lambda_2 \ell^{vgg}(\hat I_r, I_r) \label{eq:restore} \\
    L_i = & \lambda_1 [\ell_1^{msk}(\hat I_i, I_i, M\circ(A<\alpha)) + \gamma \ell_1(\hat I_i, I_i)] \nonumber \\
    & + \lambda_2 \ell^{vgg}(\hat I_i, I_i) \label{eq:imagine} \\
    L_f = & \lambda_1 [\ell_1^{msk}(\hat I, I, M) + \gamma \ell_1(\hat I, I)]  + \lambda_2 \ell^{vgg}(\hat I, I),
\end{align}
where  $\lambda_1$, $\lambda_2$, $\lambda_3$, and $\gamma$ are trade-off parameters. $\alpha$ denotes the watermark opaqueness threshold. For pixels with watermark opaqueness lower than $\alpha$, the content restoration sub-network is constrained to highlight their restoration as shown by Eq.~(\ref{eq:restore}). The content imagination sub-network is constrained to highlight the reconstruction of pixels with watermark opaqueness higher than $\alpha$ as shown by Eq.~(\ref{eq:imagine}).

The binary cross-entropy loss is adopted for constraining the predicted watermark mask $\hat M$:
\begin{equation}\label{eq:mask}
L_{m} = -\sum_{i,j}(M_{i,j}\ln(\hat{M}_{i,j}) + (1 - M_{i,j})\ln(1 - \hat{M}_{i,j})),
\end{equation}
where $M_{i,j}$ represents the value at location $(i,j)$ of $M$.
The total training loss can be formulated as: 
\begin{equation}\label{eq:total}
L = L_b+L_r+L_i+L_f+\lambda_3 L_m,
\end{equation}
where $\lambda_3$ is also a trade-off parameter.

\section{Experiments}
\subsection{Datasets and Implementation Details}

\begin{figure}[h]
\begin{center}
    \includegraphics[width=0.7\linewidth]{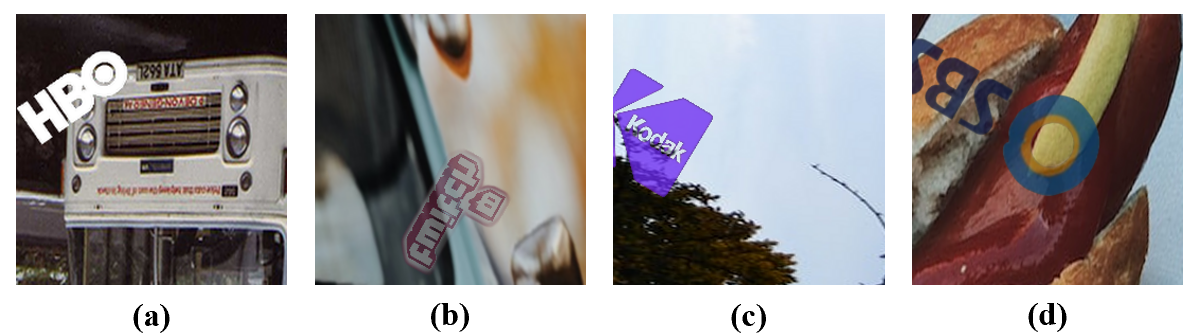}
\end{center}
   \caption{Watermarked images from our datasets: (a) totally opaque; (b) transparent; (c) moderately transparent; (d) nearly opaque.}
\label{fig10}
\end{figure}

Our experimental investigations encompass two datasets: \textit{High-opaqueness Watermarks on VOC pictures} (HWVOC) and \textit{Practical Watermarks} (PW). In these datasets, the watermarks undergo random alterations including flipping, resizing, rotation, placement and transparency adjustment prior to their composition into the background images.
Illustrative samples are showcased in Fig.~\ref{fig10}.

\textbf{HWVOC}: The background images for this dataset are collected from PASCAL VOC2012~\cite{everingham2015pascal}. Subsequently, 858 watermarks are employed for creating watermarked images, encompassing brand images of various industries such as YouTube, Amazon, and BBC. 60,000 and 2,500 watermarked images are generated for training and testing, respectively. The watermark opaqueness spans the interval (0.5, 1).

\begin{table*}[htbp]
  \centering
  \tabcolsep=3pt
  \renewcommand{\arraystretch}{1.1}
    \begin{tabular}{l|cccccc|cccccc}
    \hline
   \multirow{2}[0]{*}{Methods}       & \multicolumn{6}{c|}{HWVOC}                    & \multicolumn{6}{c}{PW} \\
          \cline{2-13}
          & PSNR & SSIM & RMSE & RMSE$_w$ & F1 & IoU (\%) & PSNR & SSIM & RMSE & RMSE$_w$ & F1 & IoU (\%) \\
          \hline
    
    WDNet & 28.22  & 0.9454  & 15.7859  & 17.8400  & 0.7123  & 59.13  & 34.57  & 0.9607  & 10.2435  & 15.4327  & 0.6517  & 52.23  \\
    \hline
    SplitNet & 38.72  & 0.9842  & 4.2730  & 15.6598  & 0.8603  & 76.96  & 40.35  & 0.9805  & 6.2317  & 12.9250  & 0.8073  & 73.36  \\
    \hline
    SLBR  & 38.26  & 0.9827  & 4.4863  & 15.2462  & 0.8379  & 74.22  & 40.26  & 0.9828  & 5.6206  & 14.7685  & 0.8207  & 75.00  \\
    \hline
    LaMa  & 37.02  & 0.9757  & 5.0132  & 20.3739  & -     & -     & 32.14  & 0.9611  & 10.9339  & 39.7764  & -     & - \\
    \hline
    RIRCI & \textbf{39.52} & \textbf{0.9855} & \textbf{3.8647} & \textbf{14.7919} & \textbf{0.8802} & \textbf{79.82} & \textbf{41.89} & \textbf{0.9866} & \textbf{4.5423} & \textbf{13.5077} & \textbf{0.8528} & \textbf{78.31} \\
    \hline
    
    \end{tabular}%
    \caption{Experimental results on HWVOC and PW datasets. The best results are in boldface.}
  \label{tab1}%
\end{table*}%
\begin{figure*}[t]
\begin{center}
    \includegraphics[width=1\linewidth]{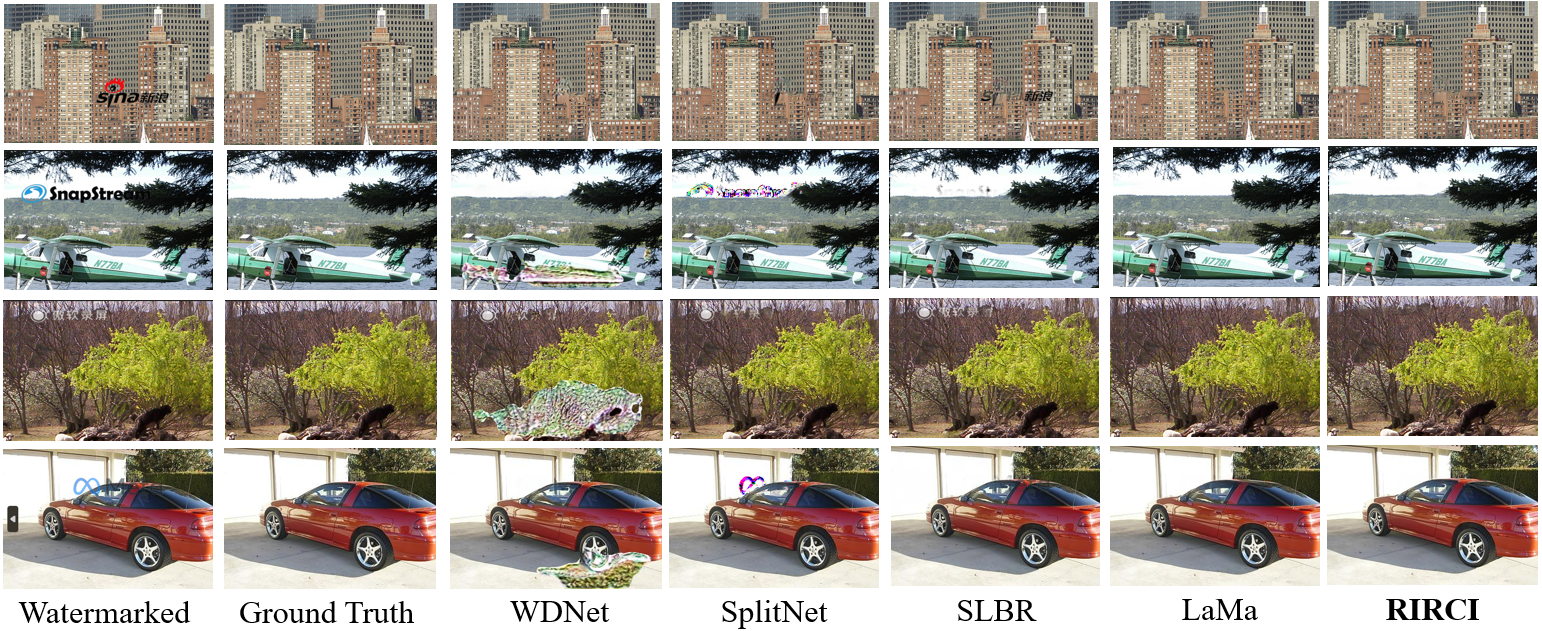}
\end{center}
   \caption{Visualization of different visible watermark removal models.}
\label{fig8}
\end{figure*}
\begin{figure*}[h]
\begin{center}
    \includegraphics[width=0.75\linewidth]{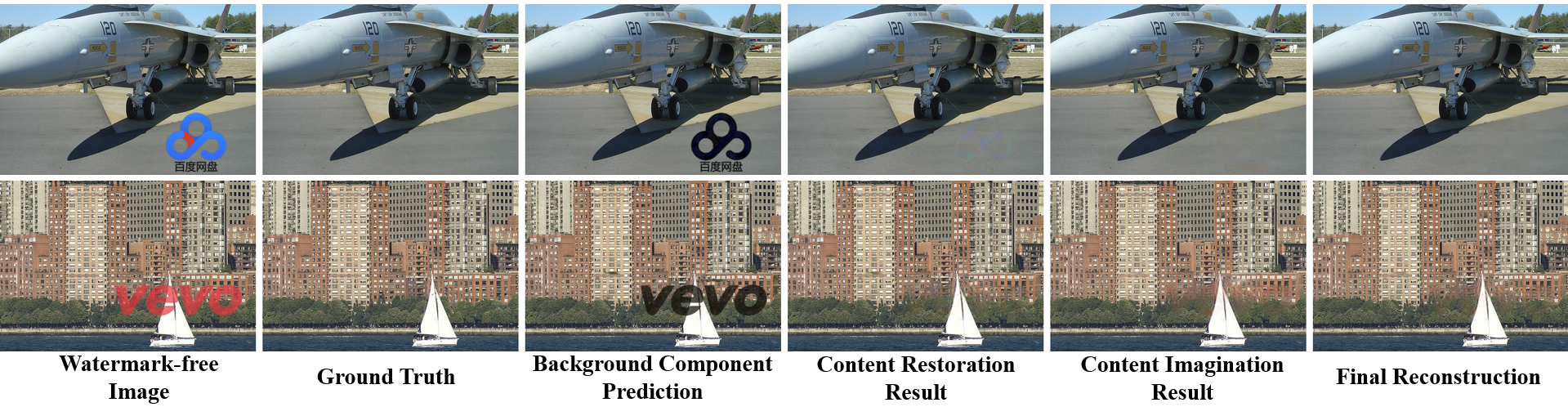}
\end{center}
   \caption{Visualization of intermediate images predicted by our method.}
\label{fig9}
\end{figure*}

\textbf{PW}: This dataset is generated from 2435 private background images. We synthesize watermarked images with analogous watermarks of HWVOC. This dataset contains 60,000 images for training and 1,045 images for testing. The watermark opaqueness interval is (0.1, 1).

All images have spatial size of $256\times 256$.
Our implementation is carried out using PyTorch \cite{paszke2019pytorch}. We train the model for 100 epochs, utilizing pretrained SLBR parameters. We adopt Adam optimizer \cite{kingma2014adam} with learning rate of 0.001, batch size of 8, $\beta_1$ of 0.9, and $\beta_2$ of 0.999. The hyper-parameters used in the training loss are: $\gamma = 1.5$, $\alpha = 0.75$, $\lambda_1=2$, $\lambda_2=1$, and $\lambda_3=3$.

\subsection{Baseline Models and Evaluation Metrics}

We train multiple visible watermark removal models using our datasets, including SLBR~\cite{liang2021visible}, SplitNet~\cite{cun2021split}, and WDNet~\cite{liu2021wdnet}. We also test the performance of LaMa~\cite{suvorov2022resolution} on our datasets, in which the ground-truth watermark mask is regarded as the inpainting hole.

We employ widely applied image restoration metrics to assess the performance of visible watermark removal models quantitatively, including Peak Signal-to-Noise Ratio (PSNR), Structural Similarity (SSIM)~\cite{wang2004image}, Root-Mean-Square distance (RMSE), and weighted Root-Mean-Square distance (RMSE$_w$). RMSE$_w$ calculates RMSE within the mask. 
The F1-score and IoU metrics are used for evaluating the accuracy of predicted watermark masks.

\subsection{Experimental Results}

Experimental results on HWVOC and PW datasets are summarized in Table~\ref{tab1}, highlighting our approach's consistent top performance across all metrics. The following sections provide an in-depth analysis of the experimental findings.

On the HWVOC dataset, our innovative dual-path predictions significantly surpass prior watermark removal models in dealing with relatively opaque watermarks. On the PW dataset, our approach's metric values highlight its universal effectiveness across watermarks with a wider range of opaqueness. 
This is further validated by Table \ref{tab11}, presenting statistical PSNR metrics on test images of PW having different opaqueness ranges.
Our proposed model distinguishes itself by the disentanglement of the watermark component exclusion and background content restoration, and the dual-branch  design in the second stage. 
It significantly improves the removal of watermarks across diverse opaqueness levels.

Visualization results are shown in Fig.~\ref{fig8}. The restoration outcomes of SLBR, WDNet, and SplitNet retain watermark remnants, introducing ambiguity due to variations in watermark transparency. In contrast, our method excels in scenarios with both opaque and transparent watermarks. Notably, LaMa's outcomes showcase unwarranted distortions, such as superfluous windows on the red building in the first row, and omission of the rearview within the car in the fourth row, while our model consistently delivers authentic restorations.

Furthermore, Fig.~\ref{fig9} displays images predictions. When encountering a image with opaque watermark (row 1), the imagination result is more similar to the ground truth. For a image with transparent watermark, the restoration result should be more applicable. Anyhow, the final reconstruction is satisfactory, which indicates the validity of our design. 

\begin{table}[t]
  \centering
  \tabcolsep=3pt
  \renewcommand{\arraystretch}{1.1}{
    \begin{tabular}{c|ccccc}
    \hline
    Opacity Range & WDNet & SplitNet & SLBR  & LaMa  & RIRCI \\
    \hline
    [0.1,0.4) & 29.04  & 35.97  & 38.02  & 33.38  & \textbf{43.21} \\
    \hline
    [0.4,0.7) & 34.55  & 40.20  & 40.50  & 32.85  & \textbf{42.47} \\
    \hline
    [0.7,1) & 35.29  & 40.99  & 40.40  & 31.56  & \textbf{41.38} \\
    \hline
    \end{tabular}%
    \caption{PSNR for test images in PW with different opacity.}
    \label{tab11}
    }
\end{table}%
\subsection{Ablation Study and Cross-dataset Validation}

\subsubsection{Design of the overall framework}
This section presents meticulously designed experiments, exploring different configurations of our methodology to confirm its robustness and efficacy. Metric results are summarized in Table~\ref{tab2}.

Observing the first row of Table~\ref{tab2}, direct prediction of the background image in stage 1 yields unsatisfactory results. As illustrated by Table~\ref{tab3}, this also hinders the accurate segmentation of the watermark, since there exists conflict between background image recovery and watermark segmentation. 
Insight from the second row of Table~\ref{tab2} underscores GLCI module's superiority over the FFC module, since GLCI introduces a more sophisticated feature propagation design of MLPs and spectral transform.
The outcomes from the third and fourth rows of Table~\ref{tab2} indicates that using a single sub-network in stage 2 performs worse than the final dual-path design. This demonstrates that the content restoration and imagination sub-networks have complementary effect in recovering the background image.
\begin{table}[t]
  \centering
  \renewcommand{\arraystretch}{1.1}
    \begin{tabular}{ccccc}
    \hline
     Variants     & PSNR & SSIM & RMSE & RMSE$_w$ \\
          \hline
    \#1 & 38.74 & 0.9840 & 4.3693 & 15.6509 \\
    \hline
    \#2 & 38.34 & 0.9836 & 4.2729 & 16.4690 \\
    \hline
    \#3 & 39.15 & 0.9849 & 4.0036 & 15.2635 \\
    \hline
    \#4 & 38.78 & 0.9834 & 4.0508 & 15.5074 \\
    \hline
    RIRCI & \textbf{39.52} & \textbf{0.9855} & \textbf{3.8647} & \textbf{14.7919} \\
    \hline
    \end{tabular}%
    \caption{Experimental results of ablation studies on HWVOC dataset. \#1: Predicting the watermark-free image instead of background component in stage 1. \#2: Using FFC in \cite{suvorov2022resolution} instead of GLCI module. \#3: Only using content restoration in stage 2. \#4: Only using content imagination in stage 2. The best results are in boldface.}
  \label{tab2}%
\end{table}%

\begin{table}[t]
  \centering
  \renewcommand{\arraystretch}{1.1}
    \begin{tabular}{ccc}
    \hline
     Variants     & F1 & IoU (\%) \\
          \hline
    \#1 & 0.8576 & 77.14 \\
    \hline
    RIRCI & \textbf{0.8802} & \textbf{79.82} \\
    \hline
    \end{tabular}%
    \caption{Performance of watermark segmentation on HWVOC dataset. \#1: Predicting the watermark-free image instead of background component in stage 1.}
  \label{tab3}%
\end{table}%
\begin{table}[h]
  \centering
  \tabcolsep=3pt
  \renewcommand{\arraystretch}{1.1}
    \begin{tabular}{ccccccc}
    \hline
     Variants     & PSNR & SSIM & RMSE & RMSE$_w$ & F1 & IoU (\%) \\
          \hline
    \#1   & 38.42  & 0.9837  & 4.3715  & 15.8798  & 0.8660  & 77.79  \\
    \hline
    \#2   & 38.79  & 0.9841  & 4.2825  & 16.0869  & 0.8624  & 77.63  \\
    \hline
    \#3   & 38.66  & 0.9843  & 4.2753  & 15.6494  & 0.8648  & 77.62  \\
    \hline
    \#4   & 39.03  & 0.9848  & 4.1280  & 15.7537  & 0.8614  & 77.33  \\
    \hline
    RIRCI & \textbf{39.52} & \textbf{0.9855} & \textbf{3.8647} & \textbf{14.7919} & \textbf{0.8802} & \textbf{79.82} \\
    \hline
    \end{tabular}%
    \caption{Experimental results of ablation studies on HWVOC dataset. \#1: Replacing GLCI by Conv 3$\times$3. \#2: Replacing scSE and spectral transform in GLCI by Conv 3$\times$3. \#3: Replacing scSE in GLCI by Conv 3$\times$3. \#4: Replacing spectral transform in GLCI by Conv 3$\times$3.}
  \label{tab4}%
\end{table}%

\subsubsection{Design of the GLCI module }
We attempt to replace GLCI module's constituent blocks with 3$\times$3 convolution layers (Conv 3$\times$3). The experimental results are summarized in Table~\ref{tab4}. 
When substituting the GLCI with Conv 3$\times$3 (\#1), there is a marked decrease in performance. This is because of the conventional convolution's inadequacy in capturing context features across varying scales. Directly combining local and global features (\#2) can be detrimental, negating the benefits of distinct feature modeling. In our method, the local-global feature transformation designs are implemented with spectral transform and scSE modules, respectively, resulting in performance enhancements (\#3 and \#4).
\begin{table}[h]
  \tabcolsep=3pt
  \renewcommand{\arraystretch}{1.1}
    \begin{tabular}{ccccccc}
    \hline
     Variants     & PSNR & SSIM & RMSE & RMSE$_w$ & F1 &IoU (\%) \\
          \hline        
    \#1   & 39.82  & 0.9828  & 5.7628  & 14.2507  & 0.8051  & 70.18  \\
    \hline
    \#2   & 37.44  & 0.9820  & 5.2627  & 19.1079  & 0.7192  & 59.07  \\
    \hline
    \end{tabular}%
    \caption{Experimental results of cross-dataset validation. \#1: Performance of our model trained with HWVOC on PW. \#2: Performance of our model trained with PW on HWVOC.}
  \label{tab5}%
\end{table}%

\subsubsection{Cross-dataset validation}
To assess our model's generalization capacity, we conduct cross-dataset validation between HWVOC and PW. 
The model trained on HWVOC is tested on PW, and vice versa. Results are summarized in Table~\ref{tab5}. 
The metric values of our method closely matches those of SLBR and SplitNet in Table~\ref{tab1}. 
This suggests our model with cross-dataset training performs comparably to previous state-of-the-art models.

\section{Conclusion}
In this paper, we introduce a novel visible watermark removal model, named \textit{Removing Interference and Recovering Content Imaginatively}. 
We first build up a watermark component exclusion model to predict the watermark mask and background component simultaneously. 
Subsequently, we set up a dual-path background restoration model which can explicitly explore both the residual background component beneath the watermark and peripheral context information from unaffected regions.
This framework helps disentangle the feature representations for removing watermarks and restoring the background content, thus mitigating the optimization conflict between the two sub-tasks.
The dual-path design is beneficial for improving the robustness of our model in coping with watermarks having diversified opaqueness.
Furthermore, we introduce a global and local context interaction module as the basic block of the background restoration model, enabling the extraction of comprehensive feature representations.
Experiment results on two large-scale datasets demonstrate the superiority of our model against existing models.

\section{Acknowledgements}
This work was supported in part by the National Natural Science Foundation of China (NO.~62376206, NO.~62003256, NO.~62322608), in part by the Shenzhen Science and Technology Program (NO.~JCYJ20220530141211024), in part by the Open Project Program of the Key Laboratory of Artificial Intelligence for Perception and Understanding, Liaoning Province (AIPU, No.~20230003), in part by Guangdong Talent Program under Grant 2021QN02X826, in part by Guangdong Key Lab of Mathematical Foundations for Artificial Intelligence, and in part by Shenzhen Science and Technology Program.

\bibliography{main}

\end{document}